\documentclass[a4paper,10pt,twoside]{cpc-hepnp}
\usepackage{CJK,upgreek,fancyhdr}
\usepackage{multicol}
\usepackage{graphicx}
\usepackage{amssymb,bm,mathrsfs,bbm,amscd}
\usepackage[tbtags]{amsmath}
\usepackage{lastpage}
\usepackage{longtable}
\usepackage{changes}
\usepackage{etoolbox}
\usepackage{setspace}
\usepackage{tabularx}
\usepackage{lineno,hyperref}
\usepackage{supertabular}
\usepackage[normalem]{ulem}
\usepackage{eqnarray}
\usepackage{booktabs}  
\usepackage{subfigure}
\begin{document}
\begin{CJK*}{GB}{gbsn}

\fancyhead[c]{\small Chinese Physics C~~~Vol. xx, No. x (202x) xxxxxx}
\fancyfoot[C]{\small 010201-\thepage}

\title{Systematic study of two-proton radioactivity half-lives within the two-potential approach with Skyrme-Hartree-Fock
\thanks{We would like to thank X. -D. Sun, J. -G. Deng, J. -L. Chen and J. -H. Cheng for useful discussion. This work is supported in part by the National Natural Science Foundation of China (Grants No. 11205083, No.11505100 and No. 11705055), the Construct Program of the Key Discipline in Hunan Province, the Research Foundation of Education Bureau of Hunan Province, China (Grant  No. 18A237), the Natural Science Foundation of Hunan Province, China (Grants No. 2018JJ2321), the Innovation Group of Nuclear and Particle Physics in USC, the Shandong Province Natural Science Foundation, China (Grant No. ZR2015AQ007), the National Innovation Training Foundation of China (Grant No. 201910555161) and the Opening Project of Cooperative Innovation Center for Nuclear Fuel Cycle Technology and Equipment, University of South China (Grant No. 2019KFZ10).}}

\author{%
\quad Xiao Pan (ÅËÏö) $^{1}$
\quad You-Tian Zou (×ÞÓÐÌð)$^{1}$
Hong-Ming Liu (ÁõºêÃú)$^{1}$
\quad Biao He (ºÎ±ë)$^{4}$\\
\quad Xiao-Hua Li (ÀîС»ª)$^{1,5,6;1)}$\email{lixiaohuaphysics@126.com}
\quad Xi-Jun Wu(Îâϲ¾ü)$^{2;2)}$\email{wuxijun1980@yahoo.cn}
\quad Zhen Zhang(ÕÅÕñ)$^{3;3)}$\email{zhangzh275@mail.sysu.edu.cn}%
}
\maketitle
\address{%
$^1$ School of Nuclear Science and Technology, University of South China, Hengyang 421001, China\\
$^2$ School of Math and Physics, University of South China, Hengyang 421001, Peoples
Republic of China\\
$^3$ Sino-French Institute of Nuclear Engineering and Technology, Sun Yat-sen University, Zhuhai 519082, China\\
$^4$ College of Physics and Electronics, Central South University, 410083 Changsha, People¡¯s Republic of China\\
$^5$ Cooperative Innovation Center for Nuclear Fuel Cycle Technology \& Equipment, University of South China, Hengyang 421001, China\\
$^6$ Key Laboratory of Low Dimensional Quantum Structures and Quantum Control, Hunan Normal University, Changsha 410081, China\\

}
\begin{abstract}
In this work, we  systematically study the two-proton($2p$)  radioactivity half-lives using  the two-potential approach while the nuclear  potential is obtained by using Skyrme-Hartree-Fock approach with the Skyrme effective interaction of {SLy8}.   For true $2p$ radioactivity($Q_{2p}$ $>$ 0 and $Q_p$ $< $0, where the $Q_p$ and  $Q_{2p}$ are
the released energy of the one-proton and two-proton radioactivity), the standard  deviation between  the  experimental half-lives and our theoretical calculations is {0.701}. In addition,  we extend this model to predict the half-lives of 15 possible $2p$  radioactivity candidates with $Q_{2p}$ $>$ 0  taken from the evaluated atomic mass table AME2016. The calculated results indicate that a clear linear relationship between the logarithmic $2p$ radioactivity half-lives $\rm{log}_{10}T_{1/2}$ and coulomb parameters [ ($Z_{d}^{0.8}$+$\emph{l}^{\,0.25}$)$Q_{2p}^{-1/2}$] considered the effect of orbital angular momentum proposed by Liu $et$  $al$ [Chin. Phys. C \textbf{45}, 024108 (2021)] is also existed.
For comparison, the generalized liquid drop model(GLDM),  the effective liquid drop model(ELDM) and Gamow-like model  are also used.  Our predicted results are consistent with the ones obtained by the other  models.
\end{abstract}


\begin{pacs}
{23.60.+e,} {21.10.Tg}
\end{pacs}

\footnotetext[0]{\hspace*{-3mm}\raisebox{0.3ex}{$\scriptstyle\copyright$}2020
Chinese Physical Society and the Institute of High Energy Physics
of the Chinese Academy of Sciences and the Institute
of Modern Physics of the Chinese Academy of Sciences and IOP Publishing Ltd}%
\begin{multicols}{2}
\section{Introduction}
In recent years, the study of exotic nuclei far from the $\beta$-stability line has became an  interesting topic in nuclear physics  with the developments of radioactive beam facilities\cite{444,777,888,999,112,114,115}. Two-proton ($2p$)  radioactivity, as an important exotic decay mode, provides a new way to obtain the nuclear structure information of rich-proton nuclei\cite{0123,0124,0125}.  In the 1960s, this decay mode  was firsrly  predicted  by Zel¡¯dovich\cite{111}  and Goldansky \cite{222,333}, independently. However due to the limitations in experiment, until 2002, the true $2p$ radioactivity($Q_{2p}$ $>$ 0 and $Q_p$ $< $0 , where the $Q_p$ and  $Q_{2p}$ are the released energy of  the proton and two-proton radioactivity)  from  $^{45}$Fe ground state was  observed at Grand acc$\acute{\rm{e}}$$\rm{l}$$\acute{\rm{e}}$rateur national dions lourds (GANIL) \cite{9} and Gesellschaft f$\ddot{\rm{u}}$r Schwerionenforschung (GSI)\cite{10}, respectively.  Later, the $2p$ radioactivity of  $^{54}$Zn, $^{48}$Ni, $^{19}$Mg and $^{67}$Kr were consecutively identified at  different  radioactive beam facilities \cite{11,12,13,14}.

 The $2p$  radioactivity process was treated as the  isotropic emission with no angular correlation or a correlated emission forming   $^2 $He-like cluster with strongly correlation from the  even-$Z$ nuclei either in the vicinity or beyond  the proton drip line\cite{9,135,147,148}. Based on the above physical mechanisms,  many 
 theoretical models have been proposed to study the  $2p$ radioactivity,  such as the direct decay model \cite{15,16,17,18,19,20,21}, the simultaneous versus sequential decay model \cite{22}, the diproton model \cite{23,24} , three-body model \cite{25,26,27,28} and so on. Moreover,  some empirical formulas can also successfully reproduce the half-lives of $2p$ radioactive nuclei  including a four-parameter empirical formula
proposed by Sreeja  $et$  $al$\cite{116} and New Geiger-Nuttall law for two-proton radioactivity  proposed by Liu $et$  $al$\cite{117}. The two-potential approach (TPA)\cite{118,119} proposed by  Gurvitz  was initially used to deal with quasi-stationary problems and has been  extended to study $\alpha$ decay, cluster  radioactivity  and  proton  radioactivity\cite{120,121,122,123,124,125,126,127,128,129,130,131,156,JG,JGD}. In our previous works,  we systematically study the proton radioactivity within the TPA while the nuclear  potential is calculated by  SHF approach\cite{132,133} denoted as TPA-SHF. The calculated results can reproduce the experimental data well. 
Since $2p$ radioactivity process may be share the similar theory of barrier penetration with proton radioactivity\cite{200,201,202,203},
whether  TPA-SHF can be extended to study the $2p$ radioactivity or not  is an interesting question.
To this end, in this work, considering the spectroscopic factor $S_{2p}$,  we extend TPA-SHF to systematically study the $2p$ radioactivity half-lives of nuclei with 4$<$Z$<$36.
For comparison, the generalized liquid drop model(GLDM)\cite{134}, the effective liquid drop model(ELDM)\cite{135} and Gamow-like models\cite{113}are also used.

This article is organized as follows. In the next section, the theoretical framework for the TPA-SHF is described in detail.   The calculated results  and discussion  are given in Section \ref{section 3}, In Section \ref{section 4}, a brief summary is given.

\section{Theoretical framework}
\label{section 2}
The $2p$ radioactivity half-life $T_\frac{1}{2}$, as an important indicator of nuclear stability, can be calculated by
\begin{equation}
T_\frac{1}{2}=\frac{ln2}{\lambda}=\frac{\hbar ln2}{\Gamma}.
\end{equation}
Here $\lambda$, $\Gamma$ and $\hbar$ are the two-proton radioactivity constant, decay width and reduced Planck constant, respectively. In the framework of  TPA\cite{118,119}, $\Gamma$ can be represented by the normalized factor \emph{F} and the penetration probability  $P$. It is expressed as
\begin{equation}
\Gamma=\frac{\hbar^2  S_{2p}F P }{4\mu},
\end{equation}
where $S_{2p} = G^2[A/ (A -2)]^{2n}\chi ^2$ denotes the  spectroscopic factor  of the $2p$ radioactivity.  It  can be obtained by the cluster overlap approximation\cite{136}.   Here  $G^2 = (2n)! / [2^{2n}(n!)^2]$ with $n \approx(3Z)^{1/3} - 1$ \cite{137} being the average principal proton oscillator quantum number \cite{138}.  $\chi ^2$ is chosen as 0.0143 according to Cui $et\ al.$ work \cite{134}. A and Z are the mass and proton number of parent nucleus, respectively.
 $F$ is the normalized factor. It can be calculated by
 \begin{equation}
F \int_{r_1}^{r_2}\frac{1}{2 k(r)}\, dr=1,
\end{equation}
 where $k(r)=\sqrt{\frac{2\mu}{\hbar^2}\left|Q_{2p}-V(r)\right|}$ is the wave number. $\mu=m_{2p}m_{d}/(m_{2p}+m_{d}) $ is the reduced mass  with $m_{2p}$ and $m_{d}$ being the mass of the emitted two protons and the residual daughter nucleus, respectively. $Q_{2p}$ is
the released energy of the  two-proton radioactivity. $V(r)$ is the total interaction potential  between the emitted two protons and daughter nucleus which will be given more in detail in the following. $r_1$, $r_2$ and the following $r_3$ are the classical turning points. They satisfy  the conditions  $V(r_{1})=V(r_2)=V(r_3)=Q_{2p}$. The penetration probability $P$ can be obtained by
\begin{equation}
P=\exp\! [-2 \int_{r_2}^{r_3} k(r)\, dr].
\end{equation}
The total interaction potential $V(r)$ is composed  by the nuclear potential $V_N(r)$,  Coulomb potential $V_C(r)$ and the centrifugal potential $V_l(r)$. It can be written as
\begin{equation}
V(r)=V_N(r)+V_C(r)+V_l(r).
\label{subeq:5}
\end{equation}

In this work, based on the assumption that the two protons spontaneously emitted from parent nuclear share momentum $\textbf{\emph{p}}$ on average  and  the nuclear interaction potential of the emitted two protons-daughter nucleus is twice of the one between the emitted proton and daughter nucleus,  we can obtained  the nuclear potential of the emitted two protons  $V_N(r)=2U_q(\rho, \rho_q, \frac{\textbf{\emph{p}}}{2})$ with SHF.  In this  model, the nuclear effective interaction is expressed as the standard Skyrme form.  It is written as \cite{140}
\begin{eqnarray}
\label{eq2}
V_{12}(\textbf{\emph{r}}_1, \textbf{\emph{r}}_2)=&t_0(1+x_0P_\sigma)\delta(\textbf{\emph{r}}_1-\textbf{\emph{r}}_2)\nonumber\\
&+ \frac{1}{2}t_1(1+x_1P_\sigma)[\textbf{\emph{P}}'^2\delta(\textbf{\emph{r}}_1-\textbf{\emph{r}}_2)+\delta(\textbf{\emph{r}}_1-\textbf{\emph{r}}_2)\textbf{\emph{P}}^2]
\nonumber\\
&+ t_2(1+x_2P_\sigma)\textbf{\emph{P}}'\cdot\delta(\textbf{\emph{r}}_1-\textbf{\emph{r}}_2)\textbf{\emph{P}}
\nonumber\\
&+ \frac{1}{6}t_3(1+x_3P_\sigma)[\rho({\frac{\textbf{\emph{r}}_1+\textbf{\emph{r}}_2}{2}})]^{\alpha}\delta(\textbf{\emph{r}}_1-\textbf{\emph{r}}_2)
\nonumber\\
&+ i {W_0} \bm{\sigma}\cdot[\textbf{\emph{P}}'\times\delta(\textbf{\emph{r}}_1-\textbf{\emph{r}}_2)\textbf{\emph{P}}], 
\end{eqnarray}
where $t_0$--$t_3$, $x_0$--$x_3$, $W_0$ and $\alpha$ are the Skyrme parameters. $\textbf{\emph{r}}_i$ (i=1, 2) is the coordinate vector of $i$-$th$ nucleon.
 $\textbf{\emph{P}}'$ and $\textbf{\emph{P}}$ are the relative momentum operator acting on the left and right. $P_\sigma$  and $\bm{\sigma}$  are the spin exchange operator and  the Pauli spin operator.  In the  SHF model, single-nucleon potential depended on  the momentum of nucleon $\textbf{\emph{p }}$ can be calculated by\cite{141}
 \begin{equation}
U_q(\rho, \rho_q, \frac{\textbf{\emph{p}}}{2})=a(\frac{\textbf{\emph{p}}}{2})^2+b, 
\label{subeq:7}
\end{equation}
where the subscript $q$ stands for proton/neutron ($q $= $p$/$n$). Total nucleonic density $\rho$ is sum of  the proton density $\rho_p$ and neutron density $ \rho_n$.  The coefficients $a$ and $b$ can be written as
\begin{eqnarray}
\label{eq4}
a=&\frac{1}{8}[t_1(x_1+2)+t_2(x_2+2)]\rho\nonumber\\
&+ \frac{1}{8}[-t_1(2x_1+1)+t_2(2x_2+1)]\rho_q, 
\end{eqnarray}

\begin{eqnarray}
\label{eq5}
b=&\frac{1}{8}[t_1(x_1+2)+t_2(x_2+2)]\frac{k^5_{f, n}+k^5_{f, p}}{5\pi^2}\nonumber\\
&+\frac{1}{8}[t_2(2x_2+1)-t_1(2x_1+1)]\frac{k^5_{f, q}}{5\pi^2}
\nonumber\\
&+ \frac{1}{2}t_0(x_0+2)\rho-\frac{1}{2}t_0(2x_0+1)\rho_q
\nonumber\\
&+ \frac{1}{24}t_3(x_3+2)(\alpha+2)\rho^{(\alpha+1)}
\nonumber\\
&-\frac{1}{24}t_3(2x_3+1)\alpha\rho^{(\alpha-1)}(\rho^2_n+\rho^2_p)
\nonumber\\
&-\frac{1}{12}t_3(2x_3+1)\rho^\alpha\rho_q. 
\end{eqnarray} 
Here  $k_{f, q}=(3\pi\rho_q)^{1/3}$ represents the Fermi momentum. The relationship among  total energy $E$ of $2p$ emission in nuclear medium, nuclear potential and Coulomb potential  can be written as
\begin{eqnarray}
E=&2U_q(\rho, \rho_q, \frac{\textbf{\emph{p}}}{2})+\frac{\textbf{\emph{p}}^2}{2m_{2p}}+V_C(r).
\label{subeq:10}
\end{eqnarray} In this work, $E$ is obtained by the corresponding  $Q_{2p}$ with $E$=[($A-$2)/$A$]$Q_{2p}$.  Based on the premise  that the total energy keeps constant when $2p$ emit from parent nuclei, using Eq.(\ref {subeq:7})  and Eq.(\ref {subeq:10}) we obtained the momentum of two emitted protons $\left|\textbf{\emph{p}}\right|$ written as
\begin{equation}
\left|\textbf{\emph{p}}\right|=\sqrt{\frac{2(E-2b-V_c(r))}{a+\frac{1}{m_{2p}}}}.
\end{equation}

The  Coulomb potential $V_C(r)$ can be obtained from a uniformly charged sphere with radius R. It is written as
 \begin{equation}
 V_C(r)=
\begin{cases}
\displaystyle \frac{Z_d Z_{2p} e^2}{2R}[3-(\frac{r}{R})^2],   r<R,
\\
\\
\displaystyle  \frac{Z_d Z_{2p} e^2}{r},   r>R,

\end{cases}
\end{equation}
where $Z_{2p}=2$ is the proton number of the two emitted protons in $2p$ radioactivity. The radius R is given by \cite{1997}
\begin{equation}
R=1.28A^{1/3}-0.76+0.8A^{-1/3}.
\end{equation}
For the last part of Eq.(\ref {subeq:5}), centrifugal potential $V_l(r)$,  we choose  the Langer modified form since $l(l + 1) \rightarrow (l + 1/2)^2$ is necessary in one-dimensional problems \cite{142}. It can be expressed as

\begin{equation}
{V_{l}(r)=\frac{\hbar^2(l+\frac{1}{2})^2}{2{\mu}r^2},}{}
\end{equation}
where $l$ is the orbital angular momentum taken away by the  two emitted protons in $2p$ radioactivity.

\section{Results and discussion}
\label{section 3} 

   In this work, we firstly calculate the $2p$ radioactivity half-lives of nuclei with 4$<$Z$<$36 using the TPA while the nuclear  potential is obtained by SHF  and compare our calculated results with the experimental data and theoretical results calculated by GLDM\cite{134}, ELDM\cite{135} and Gamow-like models\cite{113}.  For Skyrme effective interaction, there are about 120 sets current Skyrme parameters.  The SLy series parameters are widely used to describe the different nuclear reactions in various studies and the  $\alpha$ decay  since   spin-gradient term  or   a more refined two-body cent of mass correction is considered\cite{1133,1122,1144,1188,1199}. These parameters are  listed in Table 1.  As an example, we choose the Skyrme parameters of SLy8  in  this work. The  detailed calculation results  are listed in  Table 2.  In this table, the first two columns represent the two-proton emitter and the experimental released energy of $2p$  radioactivity $Q_{2p}$.  
The  experimental data of $2p$ radioactivity half-lives, the theoretical ones obtained by GLDM,  ELDM, Gamow-like and  our model in  logarithmic form  are shown in 3$-$7 columns, respectively. From  Table  2, we can see that  the  theoretical $2p$ radioactivity  half-lives calculated by our work can reproduce experimental data well. In order to intuitively survey their deviations,  we plot the difference of  $2p$ radioactivity logarithmic half-lives between the experimental data and the ones calculated by these four models (our model, GLDM, ELDM and Gamow-like) in Fig. 1.  From  this figure, we can clearly see that  all the points representing  difference are basically within $\pm$1. Especially for $^{48}$Ni of $Q_{2p}$ = 1.350 MeV and $^{54}$Zn of $Q_{2p}$ = 1.280 MeV,
 our calculated results can better reproduce the experimental data than the other models. 
 
 To obtain further insight into the well of agreement and the systematics of results,  the standard  deviation  $\sigma$  between  the theoretical values and experimental ones is used to quantify the calculated capabilities of the above four models for  $2p$ radioactivity half-lives. In this work, it  is defined as follows:
\vspace{0.3cm}
\begin{equation}
\sigma=\left[\sum_{i=1}^{\rm{n}}  \,[\rm{log_{10}T\rm^{i}_{1/2}}(\rm{expt.}) -log_{10}T\rm^{i}_{1/2}(\rm{cal.})]^2\, / \,n\right]^{1/2}.
\label{subeq:15}
\end{equation} 
Here $\rm{log_{10}T\rm^{i}_{1/2}}(\rm{expt.})$ and $\rm{log_{10}T\rm^{i}_{1/2}}(\rm{cal.})$ denote the logarithmic forms of  experimental  and calculated  $2p$ radioactivity half-lives for the $i-th$ nucleus,  respectively.  For comparison, the $\sigma$ values  of  these four models are listed in Table 4.
 From this table, we can clearly see that  the $\sigma=  {0.701}$ for this work is better than GLDM,  Gamow-like with the same  data. It indicates  our work is suitable to study $2p$ radioactivity half-lives.

In addition, as an application, we extend our model to predict the half-lives of 15 possible $2p$ radioactivity candidates with $Q_{2p}$$>$0 taken from the  evaluated atomic mass table AME2016\cite{143,144}. For comparison,  the GLDM, ELDM and Gamow-like models are also used.
The detailed results  are given in Table 3.  In this table, the first three columns represent the $2p$ radioactivity candidates,  the experimental $2p$  radioactivity released energy  $Q_{2p}$  and orbital angular momentum $l$,  respectively.  The last four columns are the theoretical values of  $2p$ radioactivity  half-lives calculated by  GLDM,  ELDM, Gamow-like and our model in logarithmic form,  respectively.  From this table, it  is clearly seen that for short-lived  $2p$ radioactivity nuclei, the order of magnitude of most predicted results calculated by our work are consistent with the ones obtained by the other  three  models. However,  for long-lived  $2p$ radioactivity  nuclei, such as  $^{49}$Ni and $^{60}$Ge,  the magnitude of our work are less  than 2-3 order to the other three models.
 In order to further clearly  compare the evaluation capabilities of those four models,  the  relationship between  the predicted results of those four models listed in Table 3 and coulomb parameters considering orbital angular momentum ( ($Z_{d}^{0.8}$+$\emph{l}^{\,0.25}$)$Q_{2p}^{-1/2}$) i.e. New Geiger-Nuttall law for two-proton radioactivity  proposed by Liu $et$  $al$\cite{117}was plot in Fig.2.  
 From this figure, we can see that the predicted results of those four models are all  linearly dependent on ($Z_{d}^{0.8}$+$\emph{l}^{\,0.25}$)$Q_{2p}^{-1/2}$  and our work can better conform to the linear relationship. 
\begin{center}
\tabcolsep=4pt
\small
\renewcommand\arraystretch{1.8}
\begin{minipage}{7.75cm}{
\small{\bf Table 4.} The standard  deviation  $\sigma$  between  the experimental data  and theoretical ones calculated by our model, GLDM, ELDM and Gamow-like model .}
\end{minipage}
\vglue5pt
\begin{tabular*}{77.5mm}{c@{\extracolsep{\fill}}ccccc}
\hline
\toprule {\rm{model }} &our model & GLDM &ELDM &Gamow-like \\
\hline
$\sigma$&0.701(10) & $0.852(10)$  & $0.531(4)$ &	$0.844(10)$\\
\hline
\end{tabular*}
\end{center}

\begin{center}
\includegraphics[width=11cm]{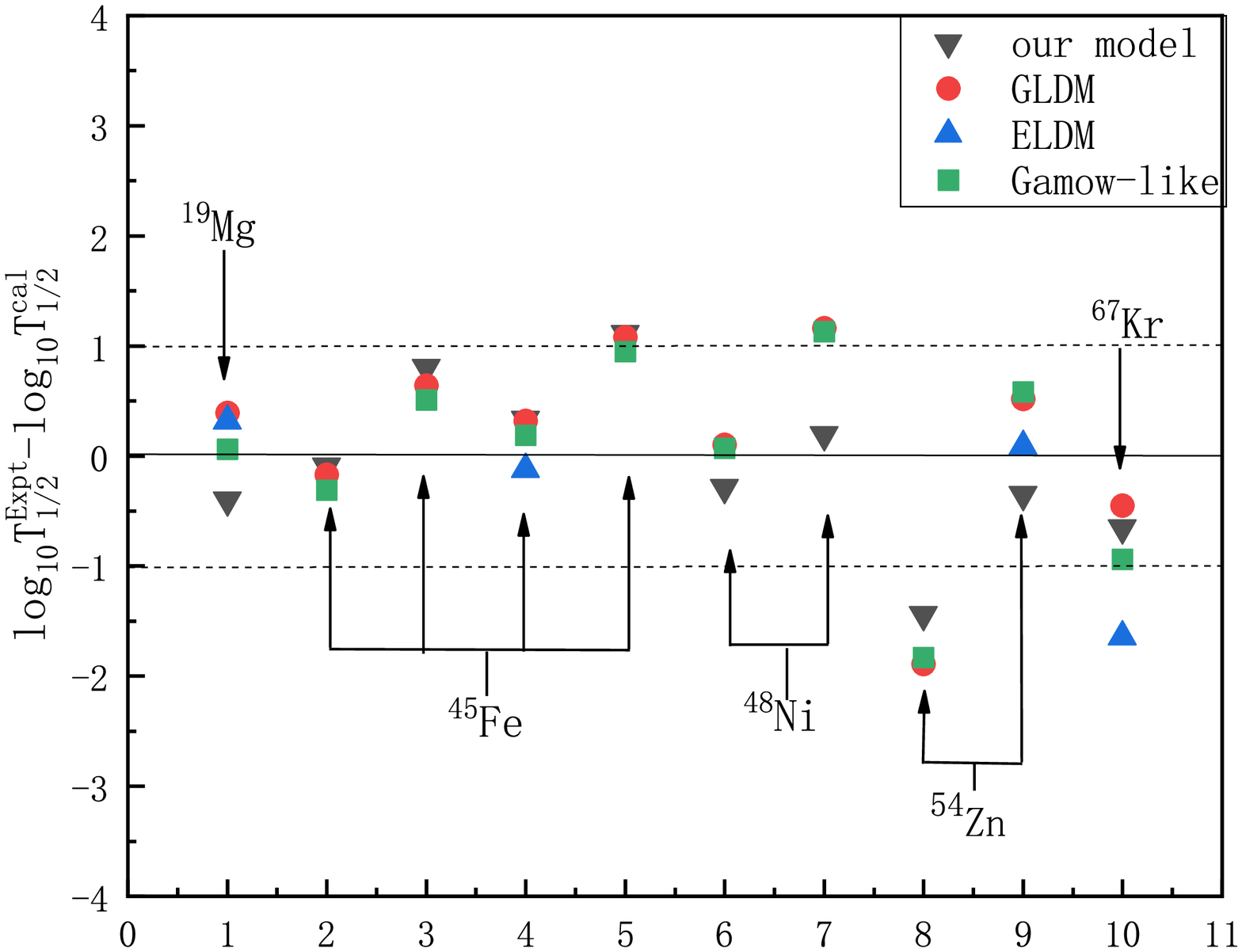}
\figcaption{(color online) The difference  between the experimental data of  $2p$ radioactivity half-lives and theoretical  onescalculated by GLDM, ELDM, Gamow-like and our model in  logarithmic form. }
\label{fig 3}
\end{center}
\end{multicols}
\begin{table*}[!hbt]
\centering
\caption{The Skyrme parameters of SLy series.  }
\label{table1}
\renewcommand\arraystretch{2.1}
\setlength{\tabcolsep}{5.8pt}
\setlength\LTleft{-13in}
\setlength\LTright{-13in plus 1 fill}
\begin{tabular}{ccccccccccc}
\hline\noalign{\smallskip}
\hline\noalign{\smallskip}
 model & $t_0$ & $t_1$ & $t_2$&   $t_3$&   $x_0$ & $x_1$ &$x_2$&$x_3$&$W_0$&$\alpha$\\
\noalign{\smallskip}\hline\noalign{\smallskip}									
Sly0\cite{1188}	&	-2486.40	&	485.20&-440.50	&	13783.0	&	0.790	&-0.500&-0.930&1.290&123.0&1/6\\
	\hline
Sly1\cite{1188} &	-2487.60	&	488.30	&	-568.90 	&	13791.0	&0.800&		-0.310&-1.000&1.290&125.0&1/6\\
Sly2	\cite{1188}&	-2484.20	&	482.20&	-290.00 	&	13763.0	&0.790&	 		-0.730&-0.780&1.280&125.0&1/6\\
Sly3\cite{1188}	&	-2481.10	&	481.00	&	-540.80	&	13731.0	&0.840&	 	-0.340&-1.000&1.360&125.0&1/6\\
Sly4\cite{1144}	&	-2488.91	&	486.82 	&	-546.39	&	13777.0	&0.834&		-0.344&-1.000&1.354&123.0&1/6\\
		
Sly5\cite{1144}	&	-2484.88	&	483.13	&	-549.40	&	13763.0	&0.778&	-0.328&-1.000&1.267&126.0&1/6\\
Sly6\cite{1144}	&	-2479.50	&	462.18	&	-448.61	&	13673.0	&0.825&-0.465&-1.000&1.355&122.0&1/6	\\
	
Sly7\cite{1144}	&	-2482.41	&	457.97	&-419.85	&13677.0 &0.846&	 	-0.511&-1.000&1.391&126.0&1/6	\\
Sly8	\cite{1188}&	-2481.40	&	480.80	&	-538.30	&	13731.0	 	&0.800&	-0.340&-1.000&1.310&125.0&1/6	\\
		
Sly9\cite{1188}	&	-2511.10	&	510.60	&	-429.80	&	13716.0	&0.800&		-0.620&-1.000&1.370&125.0&1/6\\
\noalign{\smallskip}\hline
\noalign{\smallskip}\hline
\end{tabular}
\end{table*}

\begin{table*}[!hbt]
\centering
\caption{The experimental data and  theoretical ones of $2p$ radioactivity half-lives calculated by GLDM, ELDM, Gamow-like and our model.  }
\label{table 2}
\renewcommand\arraystretch{2.1}
\setlength{\tabcolsep}{2.5pt}
\setlength\LTleft{-13in}
\setlength\LTright{-13in plus 1 fill}
\begin{tabular}{cccccccc}
\hline\noalign{\smallskip}
\hline\noalign{\smallskip}
 Nucleus & $Q_{2p}$ (MeV) &$\rm{log_{10}}{\emph{T}}_{1/2}^{\,exp}$ (s) &$\rm{log_{10}}{\emph{T}}_{1/2}^{\,GLDM}$(s)\cite{134}&  $\rm{log_{10}}{\emph{T}}_{1/2}^{\,ELDM}$(s)\cite{135}&  $\rm{log_{10}}{\emph{T}}_{1/2}^{\,Gamow-like}$(s)\cite{113} &  $\rm{log_{10}}{\emph{T}}_{1/2}^{\,our\;model}$ (s)\\
\noalign{\smallskip}\hline\noalign{\smallskip}									
$^{19}$Mg	&	0.750\cite{13}	&	$-11.40$\cite{13} &$-11.79$ 	&	$-11.72$ 	&	$-11.46$	&$-11.00$\\
	\hline
$^{45}$Fe	&	1.100\cite{10}	&	$-2.40$\cite{10} 	&	$-2.23$ 	&	--	&$-2.09$&		$-2.31$\\
	&	1.140\cite{9}	&	$-2.07$\cite{9} 	&	$-2.71$ 	&	--	&$-2.58$&	 		$-2.87$\\
	&	1.210\cite{61}	&	$-2.42$\cite{61} 	&	$-3.50 $	&	--	&$-3.37$&	 	$-3.53$\\
	&	1.154\cite{12}	&	$-2.55$\cite{12} 	&	$-2.87$ 	&	$-2.43$ 	&$-2.74$&		$-2.88$\\
		\hline
$^{48}$Ni	&	1.350\cite{12}	&	$-2.08$\cite{12} 	&	$-3.24$ 	&	--	&$-3.21$&	$-2.27$\\
	&	1.290\cite{62}	&	$-2.52$\cite{62} 	&	$-2.62 $	&	--	&$-2.59$&		$-2.23$	\\
	\hline
$^{54}$Zn	&	1.480\cite{11}	&	$-2.43$ \cite{11}	&$-2.95$ 	&$-1.32$ &$-3.01$&	 	$-2.08$	\\
	&	1.280\cite{64}	&	$-2.76$\cite{64} 	&	$-0.87$ 	&	--	 	&$-0.93$&	$-1.32$	\\
		\hline
$^{67}$Kr	&	1.690\cite{14}	&	$-1.70$\cite{14}	&	$-1.25$ 	&	$-0.06$ 	&$-0.76$&		$-1.05$\\
\noalign{\smallskip}\hline
\noalign{\smallskip}\hline
\end{tabular}
\end{table*}

\begin{center}
\includegraphics[width=15cm]{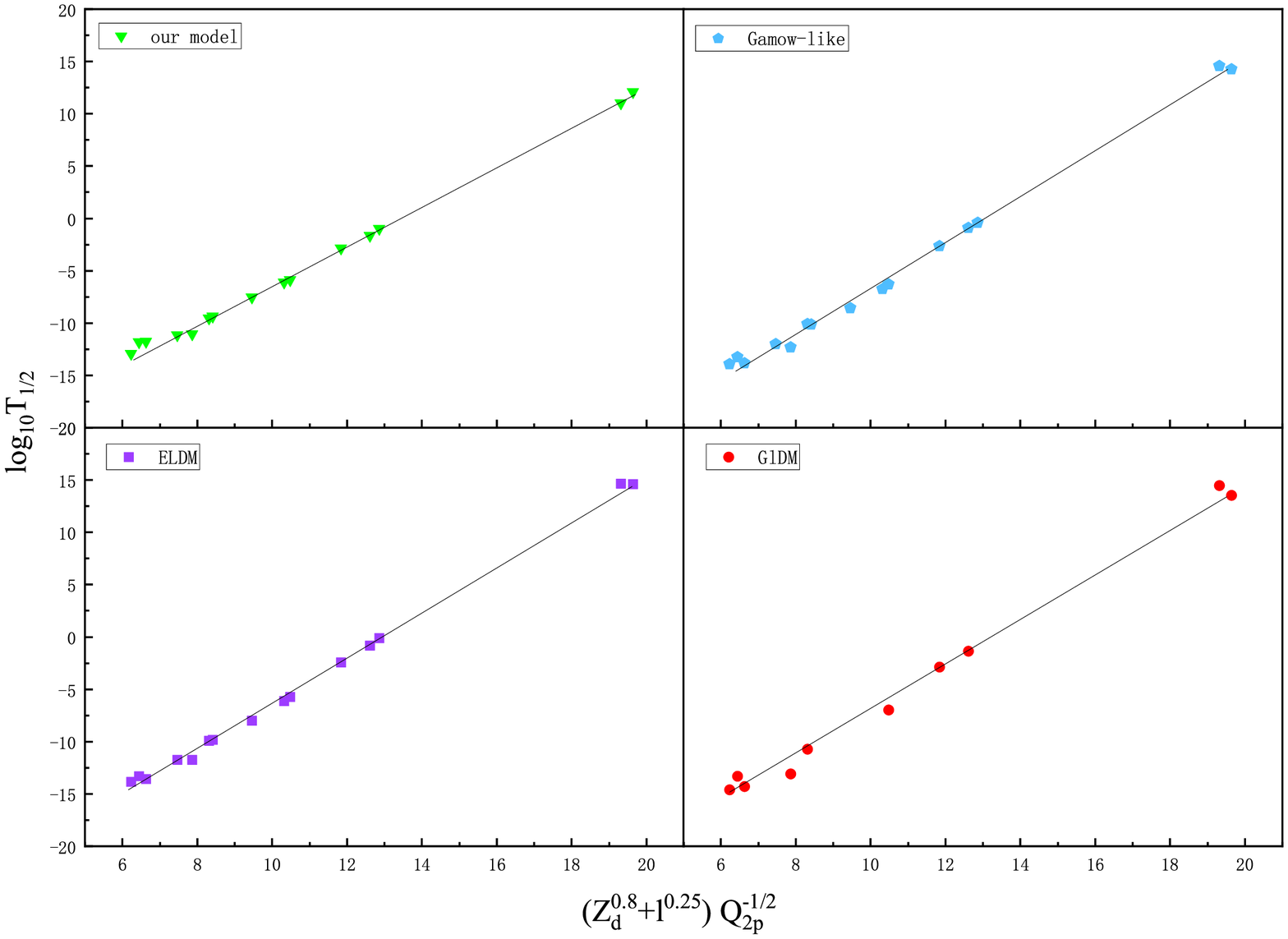}
\figcaption{(color online) The  relationship between  the predicted results of these  four models listed in Table.3 and coulomb parameters( ($Z_{d}^{0.8}$+$\emph{l}^{\,0.25}$)$Q_{2p}^{-1/2}$)  considering the effect of the orbital angular momentum  i.e. New Geiger-Nuttall law for two-proton radioactivity  proposed by Liu  $et$  $al$.}
\label{fig 3}
\end{center}

\begin{table*}[!hbt]
\centering
\caption{ Comparison of the predicted  $2p$ radioactivity half-lives using GLDM, ELDM,  Gamow-like and our model . The $2p$ radioactivity released energy $Q_{2p}$ and orbital angular momentum $l$ taken away by the two emitted protons are taken from Ref. \cite{135}.}
\label{table 4}
\renewcommand\arraystretch{2.1}
\setlength{\tabcolsep}{3pt}
\setlength\LTleft{-10in}
\setlength\LTright{-10in plus 1 fill}
\begin{tabular}{ccccccc}
\hline\noalign{\smallskip}
\hline\noalign{\smallskip}
Nucleus &$Q_{2p}$ (MeV)&$l$  &$\rm{log_{10}}{\emph{T}}_{1/2}^{\,GLDM}$ (s)\cite{134}& $\rm{log_{10}}{\emph{T}}_{1/2}^{\,ELDM}$(s)\cite{135} &  $\rm{log_{10}}{\emph{T}}_{1/2}^{\,Gamow-like}$(s)\cite{113}  & $\rm{log_{10}}{\emph{T}}_{1/2}^{\,our\;model}$ (s)\\
\noalign{\smallskip}\hline\noalign{\smallskip}
$^{22}$Si	&	1.283	&	0	&$-13.30$&$-13.32$&$-13.25$&$-11.78$\\
$^{26}$S	&	1.755	&	0	&$-14.59$&$-13.86$&$-13.92$&$-12.93 $\\
$^{34}$Ca	&	1.474	&	0	&$-10.71$&$-9.91$&$-10.10$&$-9.51$\\
$^{36}$Sc	&	1.993	&	0	&$		$&$-11.74$&$-12.00$&$-11.12$\\
$^{38}$Ti	&	2.743	&	0	&$-14.27$&$-13.56$&$-13.84$&$-11.77$\\
$^{39}$Ti	&	0.758	&	0	&$-1.34$&$-0.81$&$-0.91$&$-1.62$\\
$^{40}$V	&	1.842	&	0	&$	$&$-9.85$&$-10.15$&$-9.34$\\
$^{42}$Cr	&	1.002	&	0	&$-2.88$&$-2.43$&$-2.65$&$-2.83$\\
$^{47}$Co	&	1.042	&	0	&$		$&$-0.11$&$-0.42$&$-0.97$\\
$^{49}$Ni	&	0.492	&	0	&$	14.46 	$&$14.64$&$14.54$&$11.05$\\
$^{56}$Ga	&	2.443	&	0	&$		$&$-8.00$&$-8.57$&$-7.51$\\
$^{58}$Ge	&	3.732	&	0	&$-13.10$&$-11.74$&$-12.32$&$-11.06$\\
$^{59}$Ge	&	2.102	&	0	&$-6.97$&$-5.71$&$-6.31$&$-5.88$\\
$^{60}$Ge	&	0.631	&	0	&$	13.55 	$&$	14.62 	$&$	14.24 	$&$12.09$\\
$^{61}$As	&	2.282	&	0	&$		$&$-6.12$&$-6.76$&$-6.07$\\

\noalign{\smallskip}\hline
\noalign{\smallskip}\hline
\end{tabular}
\end{table*}
\begin{multicols}{2}
\end{multicols}
 
\begin{multicols}{2}
\section{Summary}
\label{section 4}
In the present work,  based on the  two-potential approach while the nuclear  potential is calculated by Skyrme-Hartree-Fock with the Skyrme effective interaction of { SLy8}, we  systematically study the  $2p$ radioactivity half-lives of nuclei with 4$<$Z$<$36.  The calculated results can reproduce the experimental ones  well.  In addition, we extend our model to predict the half-lives of 15 possible $2p$ radioactivity candidates with $Q_{2p}$$>$0 taken from the  evaluated atomic mass table AME2016 and compared our calculated results with the theoretical one calculated by  GLDM,  ELDM and Gamow-like models.  The predicted results of these four models are all  linearly dependent on ($Z_{d}^{0.8}$+$\emph{l}^{\,0.25}$)$Q_{2p}^{-1/2}$ i.e. New Geiger-Nuttall law for two-proton radioactivity  proposed by Liu $et$  $al$.

\end{multicols}
\vspace{-1mm}
\centerline{\rule{80mm}{0.1pt}}
\vspace{2mm}

\begin{multicols}{2}

\end{multicols}

\clearpage
\end{CJK*}
\end{document}